\documentclass{iopart}
\usepackage{epsfig}
\def\vec#1{\mathbf{#1}}

\begin{document}

\author{Astrid S. de Wijn and A. Fasolino}
\title{Relating chaos to deterministic diffusion of a molecule adsorbed on a surface}

\begin{abstract}
Chaotic internal degrees of freedom of a molecule can act as noise and affect the diffusion of the molecule on a substrate.
A separation of time scales between the fast internal dynamics and the slow motion of the centre of mass on the substrate makes it possible to directly link chaos to diffusion.
We discuss the conditions under which this is possible, and show that in simple atomistic models with pair-wise harmonic potentials, strong chaos can arise through the geometry.
Using molecular-dynamics simulations, we demonstrate that a realistic model of benzene is indeed chaotic, and that the internal chaos affects the diffusion on a graphite substrate.
\end{abstract}

\maketitle

\section{Introduction}

Diffusion is always related to noise in some way.
In the case of Brownian motion, and in most studies of diffusion, the noise is thermal in origin.
Thermal noise is, in principle, generated by a deterministic many-particle system, but noise can also be generated by smaller deterministic systems if they possess the right dynamical properties.

In the case of diffusion of large molecules adsorbed on a surface, the internal dynamics of the molecules can be such a source of noise.
Experimental \cite{longjumps} and numerical \cite{mdcarboxylicacid} evidence exists demonstrating that rotational and other internal degrees of freedom affect the diffusion.  In some cases clusters of molecules have been shown to diffuse more rapidly than single particles \cite{porphyrinsnanolet,fuscochap38}.
All this suggests that the internal degrees of freedom can play an important role.  They may act as a finite heat bath and drive the diffusion.
For this to happen, two important conditions must be met.  First of all, the internal dynamics must be chaotic, so as to produce noise.
Secondly, the internal degrees of freedom must be suitably coupled to the motion on the substrate, so that the noise can be transferred from one system to the other.  
A large body of work exists on the relationship between noise and chaos in finite systems, see for instance references \cite{noise1,noise2,bianucci,bianuccilrt,tss2,riegertnew}.

In reference~\cite{onsbball} it was shown that in a formally similar system, the bouncing-ball billiard, such a relation between internal noise and global diffusion can indeed be made.
Under certain conditions, the system exhibits a separation of time scales between slow global coordinates and the fast ``internal'' dynamics of the particle.
If the fast internal dynamics are chaotic it can be shown that they act as a source of noise for the slow global coordinates and lead to diffusion, the size of which can be estimated.
If the internal dynamics are not chaotic, there is no noise, and hence no diffusion.

In this paper we study the relation between time-scale separation, fast chaos, and diffusion of molecules on a substrate.
We describe the molecules and their interaction with the substrate with realistic, classical, atomistic Hamiltonians and show how even pair-wise harmonic potentials can lead to nonlinearities and chaotic behaviour.
We show that in benzene chaotic internal degrees of freedom do indeed lead to diffusion of the molecule on a graphite substrate, without any addition of thermal noise.

In the following section~\ref{sec:tss} we first discuss time-scale separation in adsorbed molecules, and review the case of the bouncing-ball billiard.
In section~\ref{sec:sourcesnlin} we explain how the internal degrees of freedom of a molecule can produce noise through nonlinear dynamics, which conditions must be met, and which properties of the molecule may enhance the chaos.
Finally, in section~\ref{sec:benzene}, we discuss a simple atomistic model for a benzene molecule, adsorbed on a graphite surface, and show some preliminary results, confirming that chaos in the internal degrees of freedom does indeed affect the diffusion on the substrate.

\section{Separation of time scales\label{sec:tss}}

When a molecule is adsorbed weakly on a surface, i.~e. it is not strongly distorted by the surface, the internal forces are stronger than the forces exerted by the substrate, and so the internal coordinates change more rapidly than the coordinates of the centre of mass on the substrate.
The system has two separate time scales, a long one for the motion of the centre of mass on the substrate and a shorter one for the internal degrees of freedom of the molecule.
More specifically, the equations of motion can be split into those for the slow motion of the centre of mass, described by $\vec{a}$ (containing both position and momentum), and those for the fast internal coordinates $\vec{b}$,
\begin{eqnarray}
\label{eq:tss1}
\dot{\vec{a}} = \vec{F}(\vec{a},\vec{b},t)~,\\
\dot{\vec{b}} = \frac1\epsilon \vec{G}(\vec{a},\vec{b},t)~,
\label{eq:tss2}
\end{eqnarray}
with $\vec{F}$ and $\vec{G}$ both well-behaved functions of order $1$, and the time-scale separation parameter $\epsilon$ a small number.

Time-scale separation is very common in nature and exists on many different scales and levels of complexity, from the fairly simple orbits of planets and moons to the extremely complicated interaction between the day-to-day weather and the climate.
In general, to simplify description of the slow subsystem, one wishes to eliminate the fast variables.
In practice, however, the fast variables cannot simply be removed, but must be replaced with some effective behaviour.
Several strategies exist for dealing with systems with multiple time scales.

If the fast subsystem behaves (quasi-)periodically, the coupling to the slow system can be averaged out in the treatment of the slow system\cite{averaging}.
The coupling is replaced by an effective deterministic coupling term.

If the fast subsystem is highly disordered and has a very large number of degrees of freedom, it can be approximated by an infinite heat bath \cite{vankampenadiabaticelim}.
A fast subsystem decays to equilibrium quickly, so that the coupling which affects the slow system can be approximated by equilibrium behaviour.
Using projection-operator methods, such as described in van Kampen \cite{vanKampen} or in Fick \& Sauermann \cite{FickSauermann},
it can be shown that, in general, coupling to a system consisting of many fast degrees of freedom can be approximated by stochastic forces and damping.

\subsection{Finite heat bath}

Here, we wish to consider fast subsystems which can are neither infinite nor (quasi-)periodic, but can be seen as finite heat baths.
It has been suggested that chaotic fast degrees of freedom cannot be distinguished from noise \cite{noise1,noise2,tss3}.
Therefore, we would like to replace the fast degrees of freedom by some suitably chosen noise.
If the fast system is sufficiently well-behaved and correlation in it decays exponentially, the noise should be Gaussian white noise.
In references~\cite{tss1,tss3,tss2,riegertnew} it was shown through the use of projection-operator methods, that, under certain conditions, a reduction similar to the one for large systems is also valid if the slow system is coupled to a small number of fast chaotic degrees of freedom.
Similar results were also obtained for different conditions in references~\cite{bianucci,bianuccilrt} through the use of the Zwanzig projection method, and in reference~\cite{falcionirelaxation}.  In these works the validity of linear response (or time-scale separation) is considered carefully.
In all of these results, the fast chaos acts as a finite thermodynamic heat bath and leads to stochastic driving and damping of the slow system, with clear signatures \cite{tss3,riegertnew} of the finite size of the fast system, and the finite amount of energy stored in it.

When a dynamical system explores the entire phase space for almost all initial conditions, it is called mixing.
If the fast subsystem has this property, the system's long-time behaviour is independent of the initial conditions.
If the fast subsystem also has exponential decay of correlation, then the slow coordinates can be described by a Fokker-Planck equation for their probability density.
In reference~\cite{tss2,riegertnew} explicit expressions were derived for the effective drift and diffusion in terms of averages over trajectories of the fast subsystem for fixed slow coordinates.
These expressions can be used to calculate the effective behaviour of the slow subsystem \cite{onsbball,nili}.

Due to the mixing, the expressions for the diffusion and drift are uniquely defined, and cannot depend on the history of the slow subsystem.
However, mixing is not strictly necessary.
A somewhat relaxed but more complicated condition, that only the chaotic region of the fast subsystem be mixing and deformed continuously with the slow coordinates, suffices \cite{onsbball}.
This property is much more common in physically motivated (Hamiltonian) dynamical systems with a small number of degrees of freedom than mixing itself.
 
\subsection{Time-scale separation and chaos in the bouncing-ball billiard}

In order to gain understanding of this kind of time-scale separation in physical systems, in reference~\cite{onsbball} de Wijn and Kantz studied a simpler but still physical discrete dynamical system called the bouncing-ball billiard.
It is a simple model for transport and is formally similar to transporting systems with internal degrees of freedom, such as molecules adsorbed at surfaces.

\begin{figure}
\centerline{
\epsfig{figure=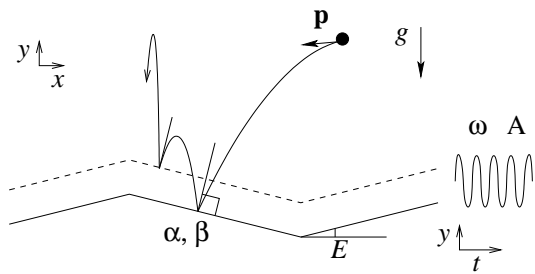,width=7cm}
}
\caption{
\label{fig:bbb}The bouncing-ball billiard with all its parameters.
The horizontal motion on the periodic profile of the driving plate, is formally similar to the motion of a particle on a substrate.
The bouncing ball itself corresponds to the internal degrees of freedom of a molecule.
}
\end{figure}

The bouncing ball billiard is a two-dimensional extension of the one-dimensional bouncing-ball problem, which has been studied in great detail both experimentally \cite{tar,1dbballexp} and theoretically \cite{1dbball1,1dbball2,1dbball3,1dbball4}.
It was first introduced by M\'aty\'as and Klages in reference~\cite{rainer}.
As shown in figure~\ref{fig:bbb}, a point particle subject to a gravitational field $g$ bounces inelastically (restitution coefficient $\alpha$ parallel to the surface and $\beta$ perpendicular to it) on a vibrating plate (frequency $\omega$ and amplitude $A$).
The vibrating plate has a symmetric tent shape, with small slope $E$.

The vertical motion of the point particle corresponds to the internal degree of freedom of the molecule, while the horizontal motion on the plate with a periodic shape corresponds to the motion of the centre of mass on the periodic substrate.
Note that the ``internal'' degrees of freedom of the bouncing ball are not Hamiltonian.
However, what is more relevant for the issues of effective diffusion is the structure of the phase space: as is the case for Hamiltonian systems with a small number of degrees of freedom the system can exhibit both chaotic and periodic behaviour (is not mixing).
One-dimensional Hamiltonian systems, on the other hand, are never chaotic.

For small slope, the bouncing-ball billiard has a time-scale separation between the horizontal and vertical motion.
This implies that if the vertical motion is chaotic, the horizontal motion should be diffusive, while if the vertical motion is periodic, the horizontal motion should be deterministic.
This was confirmed using molecular-dynamics simulations.
Due to the relative simplicity of the dynamics, the effective diffusion and drift can be evaluated (see equations~(18)--(23) of reference~\cite{onsbball}).
When compared to the numerical simulations, the results show that, in this model system, the time-scale separation theory for finite chaotic fast systems can be used to successfully predict the diffusive behaviour and approximate diffusion and drift.

\section{Sources of chaos and nonlinearity in atomistic models\label{sec:sourcesnlin}}

Through a separation of time scales, chaos in the internal degrees of freedom of a molecule can lead to diffusion on a substrate in the same way as the vertical chaos in the bouncing-ball billiard leads to diffusion in the horizontal profile of the driving plate.
In order for the molecule to generate noise and for this to contribute to the diffusion, the internal dynamics must be chaotic even without external fields.
When a system is chaotic, it is exponentially sensitive to initial conditions.
The rates at which infinitesimal perturbations of the initial conditions blow up for long times are referred to as the Lyapunov exponents.
Given the finite resolution of any observation, a trajectory with such a sensitivity becomes unpredictable for long times.
Seemingly random noise is thus generated deterministically from the infinite detail of the initial conditions.

The chaotic hypothesis, put forward by Gallavotti and Cohen \cite{gc1,gc2} states that many-particle systems, such as the ones typically studied in statistical physics, are strongly chaotic.
For systems with fewer degrees of freedom, such as the molecules studied in this paper, this assumption cannot be made.
We must first investigate how and when chaos appears in such systems.

Only nonlinear dynamical systems may be chaotic \cite{bobsboek,ott}.
In general, the nature of a trajectory in a nonlinear system depends on the initial conditions.
It is not unusual for both regular and chaotic behaviour to occur in the same system, but in different parts of the phase space.

\subsection{Atomistic models}

In principle, molecules should be treated quantum mechanically.
The study of chaos in quantum mechanical systems is a field of its own.  For more details, see, for instance, reference \cite{haakeqchaos}.
In general, it is possible to treat a quantum mechanical dynamical system classically if the energy spacings are sufficiently small compared to the total amount of energy available to excite states.

Here, we use realistic, classical potentials to describe molecules.
Many such potentials exist in the literature, but here we choose to start from the much used expressions of the tripos~5.2 force field \cite{tripos5.2}.
In this description, the bending and stretching vibrations around equilibrium positions contribute purely quadratic terms to the potential.
Most atomistic descriptions are similar, in the sense that the pair-wise potentials for bond length and angles are nearly or completely harmonic.
Naively, one would expect that pair-wise harmonic potentials cannot lead to nonlinear dynamics, but, as we will demonstrate below, the geometry of a molecule can still lead to nonlinear behaviour.

\subsection{dimer}

The model systems used here are all Hamiltonian and therefore conserve energy, as well as phase space volume and several other quantities.
Conserved quantities and symmetries of the dynamics can be used to generate perturbations which do not grow or shrink exponentially (Lyapunov exponent zero), and therefore do not contribute to chaos.
Perturbations in the centre of mass coordinates, for instance, grow at most linearly, not exponentially.

Due to the symmetries and conserved quantities, there needs to be a certain minimum number of coupled degrees of freedom before a Hamiltonian system can display chaos.
A one-dimensional dimer without external fields, for instance, cannot be chaotic, as it has four-dimensional phase space, and also four zero modes.  Three zero modes are associated with perturbations in the position and momentum of the centre of mass, and time-translation invariance.
Due to the conservation of phase-space volume, the fourth Lyapunov exponent must also be zero.

In three dimensions, a dimer has a 12-dimensional phase space of which the six components related to the centre of mass decouple, and cannot produce chaos.
The rotational symmetry generates four perturbations that cannot grow.
Finally, there is time-translation invariance, combined with the conservation of phase-space volume.
This leaves no degrees of freedom for the production of chaos.
For the possibility of chaos to exist, therefore, the molecule must consist of at least three particles.

\subsection{Nonlinearity from the geometry}

\begin{figure}
\centerline{\epsfig{figure=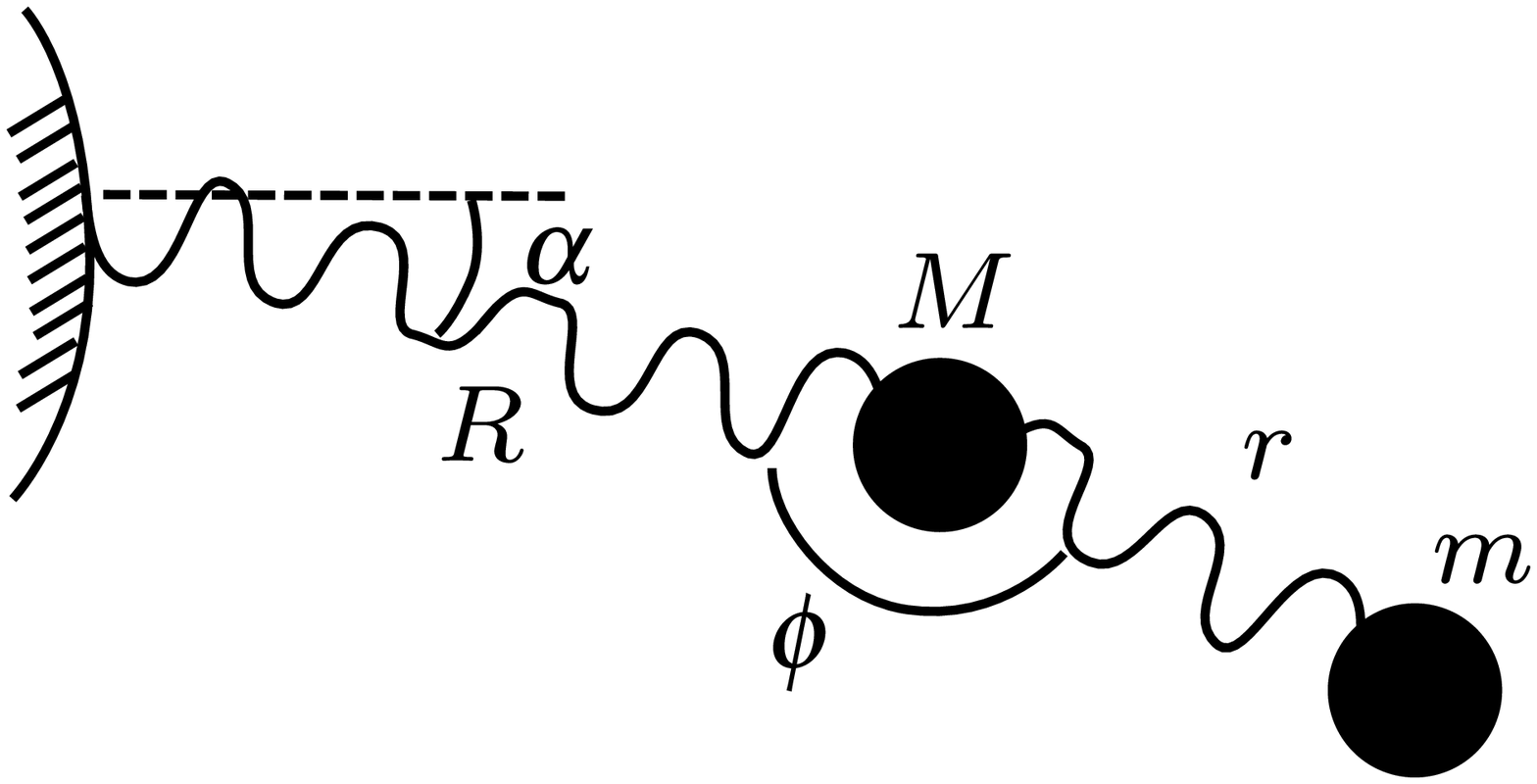, width=6cm}}
\caption{\label{fig:trimer} A simple atomistic model of a molecule with two atoms attached to a relatively heavy atom or group.}
\end{figure}

The simplest three-particle system is a trimer, such as displayed in figure~\ref{fig:trimer}.
For simplicity, we take one of the particles at the end of the trimer to be much heavier than the other two and this particle's position and orientation are kept fixed.
The two lighter atoms have masses $M$ and $m$ respectively, while the bond lengths are $R$ and $r$.
The angles $\alpha$ and $\phi$ finally describe the geometry of the bonds.
In the equilibrium configuration $R$, $r$, $\phi$, and $\alpha$ take the values $R_0$, $r_0$, $\phi_0$, and $\alpha_0$ respectively, at the global energy minimum.
Without loss of generality, $\alpha_0$ can be taken to be $0$.

As mentioned above, we use a pair-wise harmonic potential for the bending and stretching.
The typical frequencies associated with the bending are lower than those associated with the stretching, see for instance the constants in reference~\cite{tripos5.2}.
With the low frequencies come relatively large amplitudes for the same amount of energy, and so nonlinear terms in the angles or their time derivatives are larger than those in the bond lengths or their derivatives.
For the purpose of demonstrating the existence of large nonlinear terms in the dynamics, it therefore suffices to keep the bond lengths fixed.

Let $k_\phi$ and $k_\alpha$ be the spring constants of the harmonic terms for $\phi$ and $\alpha$ respectively.
The total potential energy can be written as
\begin{eqnarray}
V = \frac12 k_\phi(\phi-\phi_0)^2+ \frac12 k_\alpha \alpha^2~.
\end{eqnarray}
The kinetic energy expressed in these coordinates is more complicated,
\begin{eqnarray}
T &= &\frac12 M R^2 \dot\alpha^2\nonumber\\
&&\null + \frac12 m \left[\left(\frac{d}{dt} [R \sin\alpha + r \sin(\pi-\phi+\alpha)] \right)^2
\right.\nonumber\\ &&\phantom{[}\left.\null
+\left(\frac{d}{dt}[R \cos\alpha + r \cos(\pi-\phi+\alpha)]\right)\right]~,\\
&=&\frac12 M R^2 \dot\alpha^2 + \frac12 m \left[ (R^2+r^2) \dot\alpha^2 + r^2 \dot\phi^2
\right.\nonumber\\ &&\phantom{[}\left.\null
-2 \dot\phi\dot\alpha r^2+ 2 rR \dot\alpha(\dot\alpha -\dot\phi ) \cos(\phi-\pi) \right]~,
\end{eqnarray}
which contains harmonic, but also strongly nonlinear terms.

If the kinetic energy is expanded around the energy minimum, we see that all non-harmonic terms originate from the $\cos(\phi-\pi)$ factors.
For $\phi_0 = \pi$, these higher order terms become fourth order, whereas for $\phi_0 \neq \pi$, they are third order.
For molecules with bond angles not equal to $\pi$, the nonlinear terms are stronger and can be explored at lower energies.
The geometry produces nonlinearity, even in molecules with pair-wise harmonic potentials, and the strength of the nonlinearity depends on the bond angles.

\section{\label{sec:benzene}Diffusion of benzene on graphite}

The considerations above suggest that one should look for chaos in molecules with a sufficient number of atoms, and with somewhat complicated geometry.
Molecules such as carbon dioxide, which has the right kind of geometry, but only three atoms, may not be sufficiently nonlinear for energies corresponding to room temperature.
Larger molecules, for instance benzene, have more degrees of freedom, and are more likely to produce chaos.

\subsection{An atomistic model of benzene}

\begin{figure}
\centerline{\epsfig{figure=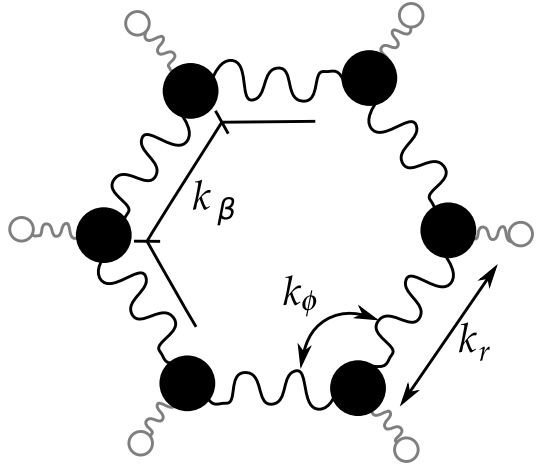,width=7cm}}
\caption{A schematic representation of the atomistic model for benzene.  The hydrogen atoms and their bonds to the carbons are difficult to treat classically and have therefore been eliminated.\label{fig:benzene}}
\end{figure}

Consider the atomistic model of benzene shown in figure~\ref{fig:benzene}.
The angles between bonds are $2\pi/3$, and the six heavy carbon atoms alone should have a sufficient number of degrees of freedom to produce chaos.
For the interaction between the atoms, we use the tripos~5.2 force field described in reference~\cite{tripos5.2}.
The bending and stretching contribute to the potential energy through quadratic terms around their energy minima.
Additionally, there is also a somewhat weaker contribution due to torsion, which depends on the positions of 4 atoms.

The hydrogen atoms in benzene are light, and the frequencies associated with various stretching and bending modes of the C-H bonds are high compared to frequencies related to interactions between heavier atoms.
Even at room temperature, the quantum states of hydrogen bonds are far apart compared to the typical available energy.
The associated modes are difficult to excite and these bonds remain in the ground state.
In other words, the hydrogen bonds do not participate in the chaotic dynamics.
If an atomistic model for benzene is to be used to investigate the chaotic dynamics, the C-H bonds must first be eliminated.

The hydrogen atoms can be averaged out using a mean-field approximation, by assuming that a hydrogen atom is, on average, in the same plane as the three nearest carbon atoms, i.~e.~the one to which it is bonded and its two neighbours.
This greatly simplifies the torsion terms.
Let $\vec{r}_i$ and $\vec{p}_i$ denote the position and momentum of the $i$-th carbon atom, ordered in such a way that $i$ and $i+1$ are neighbours.
Let $\phi_i$ be the angle between $\vec{r}_{i-1}-\vec{r}_i$ and $\vec{r}_{i+1}-\vec{r}_i$, and $\beta_i$ the torsion angle associated with the bonds between the $i-1$-st, $i$-th, $i+1$-st, and $i+2$-nd carbon atoms.
The Hamiltonian can be written as
\begin{eqnarray}
H &=& \frac1{2 (m_\mathrm{C}+m_\mathrm{H})} \sum_{i=1}^6 \vec{p}_i^2
+ \frac12 k_\phi \sum_{i=1}^6 (\phi_i-\frac23\pi)^2
\nonumber\\
&&\null
+\frac12 k_r \sum_{i=1}^6 (\|\vec{r}_{(i+1)(\mathrm{mod}~6)}-\vec{r}_{i}\|-r_0)^2
\nonumber\\
&&\null
+ \sum_{i=1}^6
k_\beta [1+\cos (2 \beta_i)]~,
\end{eqnarray}
where $k_\phi$ and $k_\beta$ are the effective bending force constant and the effective torsion constant.
The effective bending constant contains contributions from C-C-C, C-C-H, and H-C-C angles.
The effective torsion constant consists of the four torsion terms for the combinations C-C-C-C, C-C-C-H, H-C-C-C, and H-C-C-H,
\begin{eqnarray}
k_\phi = k_\mathrm{C-C-C} + \frac12 k_\mathrm{C-C-H}~,\\
k_\beta = k_\mathrm{H-C-C-H} + k_\mathrm{C-C-C-C} + 2 k_\mathrm{C-C-C-H}~.
\end{eqnarray}
The constants from reference~\cite{tripos5.2}, which will be used in this model for the rest of this paper are $r_0 = 1.47$~\AA,  $k_r = 1400~\mathrm{kcal}/(\mathrm{mol}$\AA$^2)$, $k_\phi = 158~\mathrm{kcal}/(\mathrm{rad}^2\mathrm{mol})$, and $k_\beta = 5.696~\mathrm{kcal}/\mathrm{mol}$.

\subsection{Numerical simulations of benzene}

We have performed molecular dynamics simulations of this model of benzene and numerically obtained the largest Lyapunov exponent, which quantifies the chaotic behaviour.
The system was started from random initial conditions and integrated using the velocity-Verlet algorithm.
The velocity-Verlet algorithm is symplectic and ensures that the conserved quantities in Hamiltonian systems remain conserved for finite time-step size \cite{vverlet,vverlet2}.

For a nonlinear dynamical system with a large number of dimensions it is impossible to determine numerically within reasonable time the set of initial conditions that have a specific energy.
We therefore choose the initial conditions with zero total momentum and zero angular momentum, and on the intersection of this with the constant energy shell of the linearised system, with energy corresponding to a specific temperature.
For all temperatures at which the classical description is valid, the Lyapunov exponents were found to be positive.
At energy corresponding to room temperature, 32 set of initial conditions were investigated, and all of them produced chaos.

\subsection{Benzene on a graphite substrate}

We have extended our atomistic molecular dynamics simulations of the benzene molecule described before by adding an external potential to represent the coupling to the hexagonal lattice of a graphite substrate.
We choose graphite, because of its weak interaction with adsorbed molecules.
Each carbon atom is affected separately by the potential.
The inter-atomic distance in the graphite was chosen to be equal to 1.42~\AA~and the corrugation was 25~meV.
In the direction perpendicular to the surface, the potential was harmonic with spring constant 61.2~meV/\AA$^2$.
The theoretical expressions for the effective diffusion and drift can now, in principle, be evaluated, along the lines of the calculations in \cite{onsbball,nili}.
Though the qualitative behaviour of the system is derived in this paper using very simple arguments, it should be noted that the quantitative evaluation of the theoretical expressions for the diffusion and friction is non-trivial, because the benzene molecule is far more complicated than the bouncing ball of reference~\cite{onsbball} or the hydrogen atom of reference~\cite{nili}.

\begin{figure}
\epsfig{figure=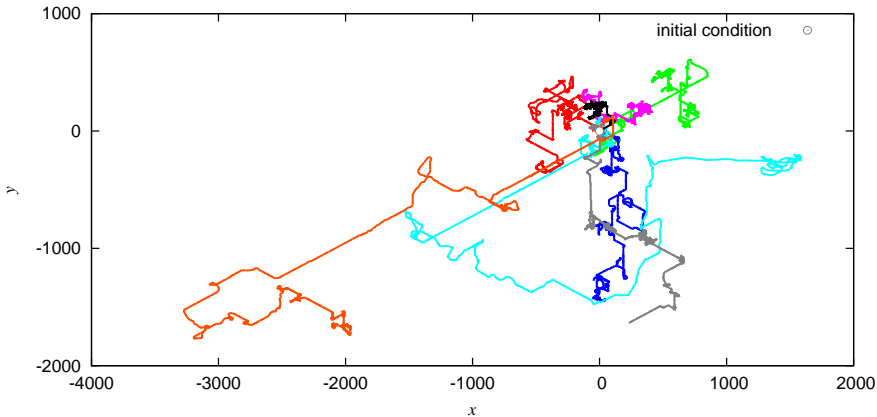, height=12cm, angle=270}
\caption{\label{fig:trajectories}The trajectories of the centres of mass of benzene molecules starting from identical initial conditions for the centre of mass, but different initial conditions for the internal degrees of freedom.  The initial condition is $(x,y) =(0,0), (p_x,p_y) = (0,0)$.  Distances are in $\mathrm{a.u.}$ ($a_0 = 0.529177$\AA).
The motion on the substrate is diffusive, and, for sufficiently long trajectories, the diffusion appears to be normal.
}
\end{figure}

Initial conditions were again chosen randomly, with the energy of the linearised system corresponding to room temperature, and with zero total momentum.
The model molecule was placed on a substrate in an energy minimum, and the system was integrated with time steps of size of 1.0~a.u=0.0242~fs.

The motion of the centre of mass is displayed in figure~\ref{fig:trajectories} for eight different sets of initial conditions.
All eight systems have a positive largest Lyapunov exponent.
Though the initial conditions for the centre of mass are identical in all cases, the eight resulting trajectories are different.
The chaotic internal dynamics therefore cause diffusion of the centre of mass on the substrate.

Analysis of the time-dependence of the mean square displacement suggests that the diffusion is normal.
However, the diffusion sometimes exhibits long jumps, which produces a dependence of the mean square displacement on time which only becomes linear for long times and for two trajectories displayed in the figure~\ref{fig:trajectories} the analysis is inconclusive.
Note that in a system which is also subject to thermal noise, extremely long jumps could be more rare.

\section{Conclusion}

The internal degrees of freedom of molecules can be related to the diffusion of their centres of mass on a substrate.
If the internal degrees of freedom are chaotic, they can be seen as a finite heat bath which generates noise that leads to diffusion of the centre of mass.
This enhances the diffusion of a molecule with internal degrees of freedom on a substrate compared to diffusion with purely thermal origin.

We have demonstrated in this paper that atomistic models of molecules with pair-wise harmonic potentials can produce chaotic behaviour.
In a realistic model of benzene, chaos occurs at energies comparable to realistic temperatures.
We have shown that the time-scale separation combined with internal chaos of the benzene molecule implies that the benzene molecule should diffuse on a sufficiently weak substrate, such as graphite without any external sources of noise.
The preliminary numerical studies of benzene on graphite presented here confirm this.

These results raise the possibility that the time-scale separation between the internal dynamics and the centre-of-mass motion could enhance the diffusion of heavy molecules in experiments.
Within the frame-work of the time-scale separation theory, theoretical expressions exist for the effective diffusion of such molecules due to their internal degrees of freedom.
However, evaluating these expressions is not a trivial exercise, as even the simple model of the benzene molecule on a graphite surface described in this paper is more complicated than the systems for which they have been evaluated in the past.
Additionally, the interplay between the noise from the finite heat bath and the thermal noise from the environment will have to be investigated further.

\section*{Acknowledgements}
ASW would like to thank NWO (The Netherlands Organisation for Scientific Research) for financial support through the VENI program.

\end{document}